\documentclass[usegraphicx,useAMS,usenatbib]{mn2e}
\usepackage{times}
\usepackage{amssymb}
\usepackage{amsmath}
\usepackage{graphicx}
\usepackage{euscript}
\usepackage{rotating}
\usepackage{epsfig}
\usepackage{mathptmx}
\usepackage{amsbsy}

\def\km{{\rm\thinspace km}}

\def\s{{\rm\thinspace s}}

\def\kmps{\hbox{$\km\s^{-1}\,$}}

\begin{document}
\include{defn} \title[O{\sc iii} emitting galaxies around the radio galaxy MRC\,0316-257]{[O{\sc iii}] emitters in the field of the 
MRC\,0316-257 protocluster \\}
\author[ F. Maschietto et al.]
{F. Maschietto$^{1}$\thanks{E-mail:maschietto@strw.leidenuniv.nl}, N.A. Hatch$^{1}$, B.P. Venemans$^{2}$,  H.J.A. R$\ddot{\rm o}$ttgering$^{1}$, G.K. Miley$^{1}$, R.A. Overzier$^{3}$,
\newauthor
M.A. Dopita$^{4}$, P.R. Eisenhardt$^{5}$, J.D. Kurk$^{6}$, G.R. Meurer$^{7}$, L. Pentericci$^{8}$, P. Rosati$^{9}$,
\newauthor
S.A. Stanford$^{10,11}$,W. van Breugel$^{11,12}$, A.W. Zirm$^{1,7}$\\
$^{1}$Leiden Observatory, Leiden University, Niels Bohrweg 2, 2333 CA Leiden, The Netherlands\\
$^{2}$Institute of Astronomy, Madingley Road, Cambridge CB3 OAH, United Kingdom\\
$^{3}$Max-Planck-Institut f\"ur Astrophysik, Karl-Schwarzschild-Str. 1, D-85741 Garching, Germany\\
$^{4}$Research School of Astronomy \& Astrophysics, Australian National University, Cotter Rd., Weston Creek ACT 2611, Australia\\
$^{5}$Jet Propulsion Laboratory, California Institute of Technology, MS169-327, 4800 Oak Grove Drive, Pasadena, CA 91109, USA\\
$^{6}$Max-Planck-Institut f\"ur Astronomie, D-69117 Heidelberg, Germany\\
$^{7}$Department of Physics and Astronomy, Johns Hopkins University, Baltimore, MD 21218, USA\\
$^{8}$ INAF--Osservatorio Astronomico di Roma, via Frascati 33, 00040 Monteporzio, Italy\\
$^{9}$European Southern Observatory, Karl-Schwarzschild-Strasse2, D-85748, Garching, Germany\\
$^{10}$University of California, Davis, CA 95616, USA\\
$^{11}$Institute of Geophysics and Planetary Physics, Lawrence Livermore National Laboratory, Livermore, CA 94551, USA\\
$^{12}$University of California, Merced, PO Box 2039, Merced, CA 95344, USA}

\maketitle

\label{firstpage}

\begin{abstract}

\noindent \citet{Venemans2005} found evidence for an overdensity of Ly$\alpha$ emission line galaxies associated with the radio galaxy MRC 0316--257 at $z=3.13$ indicating the presence of a massive protocluster. Here, we present the results of a search for additional star-forming galaxies and AGN within the protocluster.
Narrow-band infrared imaging was used to select candidate [O{\sc iii}] emitters in a 1.1$\times$1.1 Mpc$^2$ region around the radio galaxy. Thirteen candidates have been detected. Four of these are among the previously confirmed sample of Ly$\alpha$ galaxies, and an additional three have been confirmed through follow-up infrared spectroscopy. The three newly confirmed objects lie within a few hundred km s$^{-1}$ of each other, but are blueshifted with respect to the radio galaxy and Ly$\alpha$ emitters by $\sim2100$ km s$^{-1}$. Although the sample is currently small, our results indicate that the radio--selected protocluster is forming at the centre of a larger, $\sim60$ co-moving Mpc super-structure. 
On the basis of an HST/ACS imaging study we calculate dust-corrected star-formation rates and investigate morphologies and sizes of the [O{\sc iii}] candidate emitters. From a comparison of the star formation rate derived from UV-continuum and [O{\sc iii}] emission, we conclude that at least two of the [O{\sc iii}] galaxies harbour an AGN which ionized the O$^{+}$ gas.
\end{abstract}


\section{Introduction}
To understand the formation and evolution of galaxy clusters, it is desirable to find and study their high redshift progenitors. Although galaxy clusters have been found out to a redshift of $z=1.5$ \citep{Mullis2005,Stanford2005}, their higher redshift progenitors are sparse and difficult to find. A successful technique for finding more distant structures that by--passes the need for surveying very large areas of the sky is to search for emission-line galaxies in the neighbourhood of luminous high--redshift radio galaxies (HzRGs) using narrow-band imaging.  Multiwavelength studies of HzRGs have resulted in strong evidence that they are massive forming galaxies  \citep[e.g.][]{Seymour2007,Villar-Martin2006}  and are frequently associated with overdensities of emission-line galaxies \citep{Venemans2007}. These overdense regions in the early universe are the likely progenitors of local galaxy clusters or groups and are termed ''protoclusters". PROCESS (PROtoCluster Evolution Systematic Study) is a project designed to use a few key radio-selected protoclusters with $2 \le z \le 5$ to investigate the formation of and evolution of various populations of galaxies in dense environments \citep{Overzier2006,Overzier2007,Overzier2008,Venemans2005,Venemans2007}. 

This article presents observations of the protocluster surrounding the PROCESS radio galaxy MRC\,0316-257. The associated $1.5 \ {\rm Jy}$ radio source was listed in the 408 MHz Molonglo Reference Catalogue \citep{Large1981} and was optically identified with a galaxy at $z=3.13$ by \citet{McCarthy1990}. \citet{LeFevre1996} spectroscopically confirmed two Ly$\alpha$ emitting companions to the HzRG, indicating that the radio galaxy is located in a dense environment.  Recently, \citet{Venemans2005,Venemans2007} confirmed 31 Ly$\alpha$ emitters at a similar redshift of MRC\,0316-257. The corresponding overdensity is approximately 3.3 times the galaxy field density at this redshift. The protocluster redshift of $z \sim 3.13$ corresponds to an epoch when both the cosmic star formation rate and the quasar luminosity function were at their peak, indicating that this is a key epoch for studying the evolution of different populations of galaxies. 
We have identified and studied additional galaxies in the MRC\,0316-257 protocluster on the basis of their redshifted [O{\sc iii}] emission. The observations that we shall discuss here consist of infrared imaging and spectroscopy with ESO's Very Large Telescope (VLT) and deep optical imaging with the Advanced Camera for Surveys (ACS) on the Hubble Space Telescope (HST). 

Section 2 of this article is an outline of the observations and the data reduction. In Section 3 we present results from the VLT search programme and the deep ACS images. Corrected star formation rates derived from the UV fluxes are used to discriminate between star-forming galaxies and obscured AGNs and the morphologies and sizes of the candidate emitters are discussed. The implications of our results for the space density of 
[O{\sc iii}] emitting galaxies and the origin of the [O{\sc iii}] emission are discussed in Section 4 and the conclusions of the article are presented in Section 5.
We assume  a flat cosmology  with H$_0$=71 [km sec$^{-1}$ Mpc$^{-1}$] and $\Omega_{\rm m}$=0.27 \citep{Spergel2003}. At the distance of MRC\,0316-257 an angular scale of 1\,arcsec corresponds to a projected linear scale of 7.73\,kpc. All magnitudes are given in the AB system \citep{Oke1974}.

\section{OBSERVATIONS DATA REDUCTION AND SAMPLE SELECTION}

\subsection{Infrared Narrow--band Imaging and Selection of [O{\sc iii}]--Emitting Candidates}

To search for [O{\sc iii}] emitting galaxies near MRC 0316--257, narrow- and
broad-band imaging were carried out between 2003 November and 2004
October in service mode with ISAAC \citep{Moorwood1998} on the 8.2 m ESO VLT Antu (UT1). 
The pointing was chosen to include the radio galaxy and as
many of the confirmed Ly$\alpha$ emitters from \citet{Venemans2005} as possible. 

The ISAAC narrow--band filter used was NB\_2.07, with a central
wavelength ($\lambda_c$) of 2.07 $\mu$m and a width
($\lambda_{\mathrm{FWHM}}$) of 0.026 $\mu$m. This is
sensitive to [O{\sc iii}] emission at redshifts of $z\sim3.09$--3.16. 
Broad-band images were obtained to provide an ''off-band" measurement 
and the magnitude and slope of the continuum emission from candidate line
emitters. These were taken in the filters $J$ ($\lambda_c=1.25 \mu$m and
$\lambda_{\mathrm{FWHM}}=0.29 \mu$m) and $K_s$ ($\lambda_c=2.16 \mu$m
and $\lambda_{\mathrm{FWHM}}=0.27 \mu$m). The detector was an Hawaii
array with 1024$\times$1024 pixels. The pixel scale was 0.148
arcsec\,pixel$^{-1}$ and the field of view 152$\times$152
arcsec$^2$. The images were taken using dithered exposures of length 150, 75,
and  183 seconds in $J$, $K_s$ and the narrow-band, each with
sub-integrations of 30, 15, and 61 seconds to avoid over-exposure of the
background. The total exposure times were 5.3 hours each in $J$ and
$K_s$, and 6.9 hours in the narrow-band.
Standard stars were observed for the photometric calibration. The
observed standards include FS6, FS11, and P565C taken from the
UKIRT Faint Standards catalog \citep{Hawarden2001} and S363D from
the LCO/Palomar NICMOS Photometric Standards list \citep{Persson1998}.

The images were reduced using the Experimental Deep Infrared Mosaicing Software (XDIMSUM\footnote{http://iraf.noao.edu/iraf/ftp/iraf/extern/xdimsum/xdimsum.readme}). 
The effective area of the reduced images was
139$\times$139 arcsec$^2$, and the seeing, as measured from bright
stars in the field was 0.45 arcsec in the NB\_2.07 image and 0.55
arcsec in the $J$ and $K_s$ images. The magnitude zero-point of the
broad-band images obtained from the various standard star observations
has an accuracy of 0.01 mag ($J$) and 0.03 mag ($K_s$). The zeropoint
of the narrow-band image was derived using 52 objects that were
detected at a signal-to-noise of at least 20 in both $J$ and
$K_s$. Narrow-band magnitudes were computed for the 52 objects assuming
a power--law spectral energy distribution and with the associated counts
in the narrow-band image the zeropoint was derived with an accuracy of
0.01 mag.
The 5 $\sigma$ limiting magnitudes per square arcsecond were 23.9
(NB2\_2.07), 25.6 ($J$) and 24.7 ($K_s$).
Objects in the images were extracted using the program SExtractor
\citep*{Bertin1996}. The NB\_2.07 image was used as the detection image,
and aperture photometry was subsequently performed on both
the narrow- and broad-band images. For detection, objects were required to
have a signal-to-noise of $>5$ in the NB\_2.07 image. The colours of
the detected objects were measured in circular apertures, while the
``total'' flux was measured in an elliptical aperture. A
total of 143 objects having a signal-to-noise of at least 5 were detected in the narrow-band image .
Following \citet{Kurk2004}, we selected objects with a rest-frame
equivalent width {\em EW}$_0 > 50$ \AA\ and a significance $\Sigma
\equiv$ {\em EW}$_0$/$\Delta${\em EW}$_0$ $> 3$ as candidate
[O{\sc iii}] emitters (see \citet{Venemans2005} for more details on the object detection and photometry) and 
on how {\em EW}$_0$ and $\Delta${\em EW}$_0$ are
computed. Each candidate was inspected visually in order to remove
spurious sources. This resulted in a list of 17 candidate [O{\sc iii}]
emitting galaxies. 
To remove foreground galaxies with emission lines that fall
into the NB\_2.07 filter, we measured the colours of the candidates on
archival $u'$, $V$ and $I$ images \citep{Venemans2005, Venemans2007}. Three candidates had $u'-V$ colors of $<1.2$ and are likely
foreground objects. Two of the candidates lie very close together and after inspection of the galaxy morphologies were determined
to be a single merging object.
The spatial distribution of the remaining 13
candidate [O{\sc iii}] emitters is plotted in Fig.\ 1.  One of the candidates
is the radio galaxy (ID 1), four are spectroscopically confirmed
Ly$\alpha$ emitters (ID 3, 6, 7, 11), six are likely Lyman break galaxies
(LBGs, defined as objects with $u'-V > 1.6$ and $V-I<0.6$; ID 2, 3, 4,
7, 8, 10) and two satisfy the selection criteria for distant red galaxies (DRGs, $J-K_s > 1.4$;
ID 3, 5). For an overview see Table 1.

\subsection{ACS Optical Imaging}

We obtained deep images in $r_{625}$ and $I_{814}$ filters with the Advanced Camera for Surveys (ACS; \citealt{Ford1998}) on the HST. The MRC 0316--257 protocluster was observed with the ACS during 2004 December 14--31 and 2005 January 2--21 in two $3.4\arcmin\times3.4\arcmin$ ACS fields that overlapped by $\sim$1\arcmin\ in a region that includes the radio galaxy.
The exposure time per pointing was 22,410s in $I_{814}$ (10 orbits per pointing) and 11,505s in $r_{625}$ (5 orbits per pointing).
The $I_{814}$ images were combined with a 6,300s exposure (3 orbits) taken with ACS on 2002 July 18.
All observations were processed through the ACS GTO pipeline \citep{Blakeslee2003} to produce registered, cosmic-ray rejected images.
The extinctions in the direction of the radio galaxy are 0.04 and 0.03 mag at wavelengths corresponding to the $r_{625}$ and $I_{814}$ filters, respectively. In its deepest part (about 1/5 of the image), the 2$\sigma$ depths are 29.1 ($r_{625}$) and 29.6 AB mag ($I_{814}$), measured in a square $0\farcs45$ diameter aperture. In its shallowest outer part (about 2/5 of the image), the $2\sigma$ depths are 28.5 ($r_{625}$) and 29.3 ($I_{814}$ ) mag. The r$_{625}$ and I$_{814}$ filter bands correspond to a central wavelength of $\sim$ 1513 $\rm{\AA}$ and 1937 $\rm{\AA}$ in the rest-frame respectively.
We determined the rest-frame UV magnitudes and the relative RMS errors of the [O{\sc iii}] candidates from the ACS images using SExtractor, cross-checking the results using both circular and elliptical apertures with the IDL routine ATV.
All the 13 [O{\sc iii}] candidate emitters were detected in both the UV filters with the faintest having a I$_{814}$ magnitude of 26.2.

\subsection{Infrared Spectroscopy}
\begin{figure}
\includegraphics[width=1.0\columnwidth, angle=0]{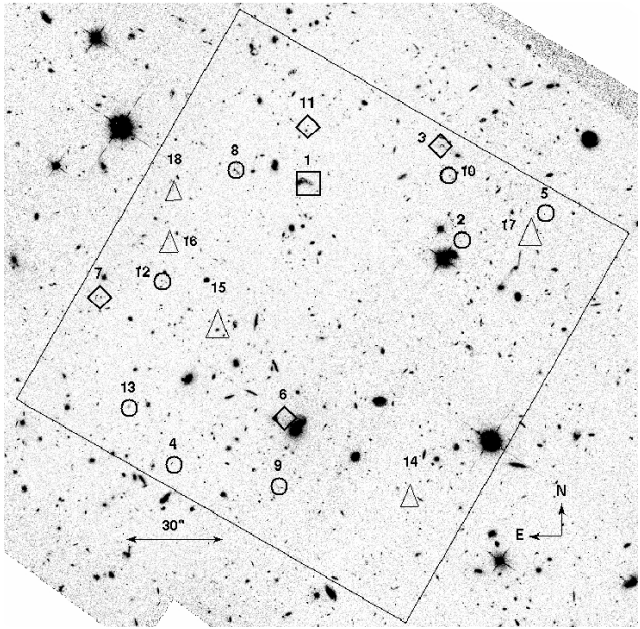}
\caption{The ACS I$_{814}$ image with the [O{\sc iii}] candidate emitting galaxies marked. The square  symbol denotes the radio galaxy, the circles mark the [O{\sc iii}] candidates, the diamonds identify the [O{\sc iii}] candidate emitters which are also confirmed Ly$\alpha$ emitters, the triangles indicate the Ly$\alpha$ galaxies with no detected [O{
\sc iii}] emission. The large box indicates the ISAAC field-of-view. \label{all_galaxies}
\label{map}}
\end{figure}

Infrared spectra of four candidate [O{\sc iii}] emitters were obtained with ISAAC \citep{Moorwood1999} at the ESO VLT Antu (UT1) on the nights of August 31, September 3 and November 23, 2006. The observations used the short wavelength camera of ISAAC and the medium resolution grating, resulting in a pixel scale of $0.148''$ and a spectral dispersion of 1.23\AA. The observations were carried out under variable optical seeing conditions, which were typically between $0.8''-1''$, except for the final part of the night of September 3, where the  seeing was $\sim0.33''$.  The $1''\times120''$ slit was employed  in all cases, except for a 90\,min observation on the night of September 3, when the slit width was changed to $0.6''$. The $1''\times120''$ slit resulted in a resolution of  7.4\AA\ full width at half maximum (FWHM; or $R \sim$ 2600). 

The target galaxies were acquired using the imaging mode of ISAAC. The [O{\sc iii}] galaxies are faint so the slit was first centered on a nearby bright star with  $17.6 \le K_{AB} \le 19.8$ and then offset to the target galaxies.  High precision is required for determining the co-ordinates of the [O{\sc iii}] emitters. However, the reduced ISAAC NB\_2.07, $J$- and $K_S$-band images that were used for the candidate selection have poor astrometric accuracy due to the low number of reference stars in the field. Therefore a $I$-band FORS image of the field,  earlier obtained \citep{Venemans2005} was used to determine the required offsets.

Four candidate [O{\sc iii}] emitters were observed, galaxies identified as 2, 4, 5 and 9 in Fig.~\ref{map}. The exposure time for candidates ID 5 and 2 was 19\,800\,s and for ID 4 and 9 was 18\,000\,s. We employed the classical ABBA sequence to remove the sky background and sky emission lines. The exposure time for each integration was either 600 or 900 seconds.
All observations were carried out at airmass smaller than 1.7. A set of Xeon and Argon arc spectra were taken during each observing night for wavelength calibration.

The data were reduced using the ISAAC data reduction software package {\sc eclipse}, which is available from ESO. The data were flat-fielded and the arcs were used to model and remove the slit curvature, and wavelength calibrate the data. The arcs from the night of  2006 November 23 contained 14 detected arclines and the standard deviation of the fit was 0.2\AA, whereas on the nights of 2006-08-31 and 2006-09-03 there were 12 and 11 lines detected with a standard deviation of 0.4\AA\ and 0.5\AA\ respectively. The wavelength of the skylines were measured to check the arc calibration. No deviations greater than 0.6\AA\ were found. The sky background and night sky emission lines were removed by the ABBA dither sequence, however a feature at 2.041$\mu$m had prominent residuals and has been masked out in the spectra shown in Fig.~\ref{spectra}. 

\section{Results}
\subsection{Redshift Distribution}
\begin{figure}
\includegraphics[width=1\columnwidth]{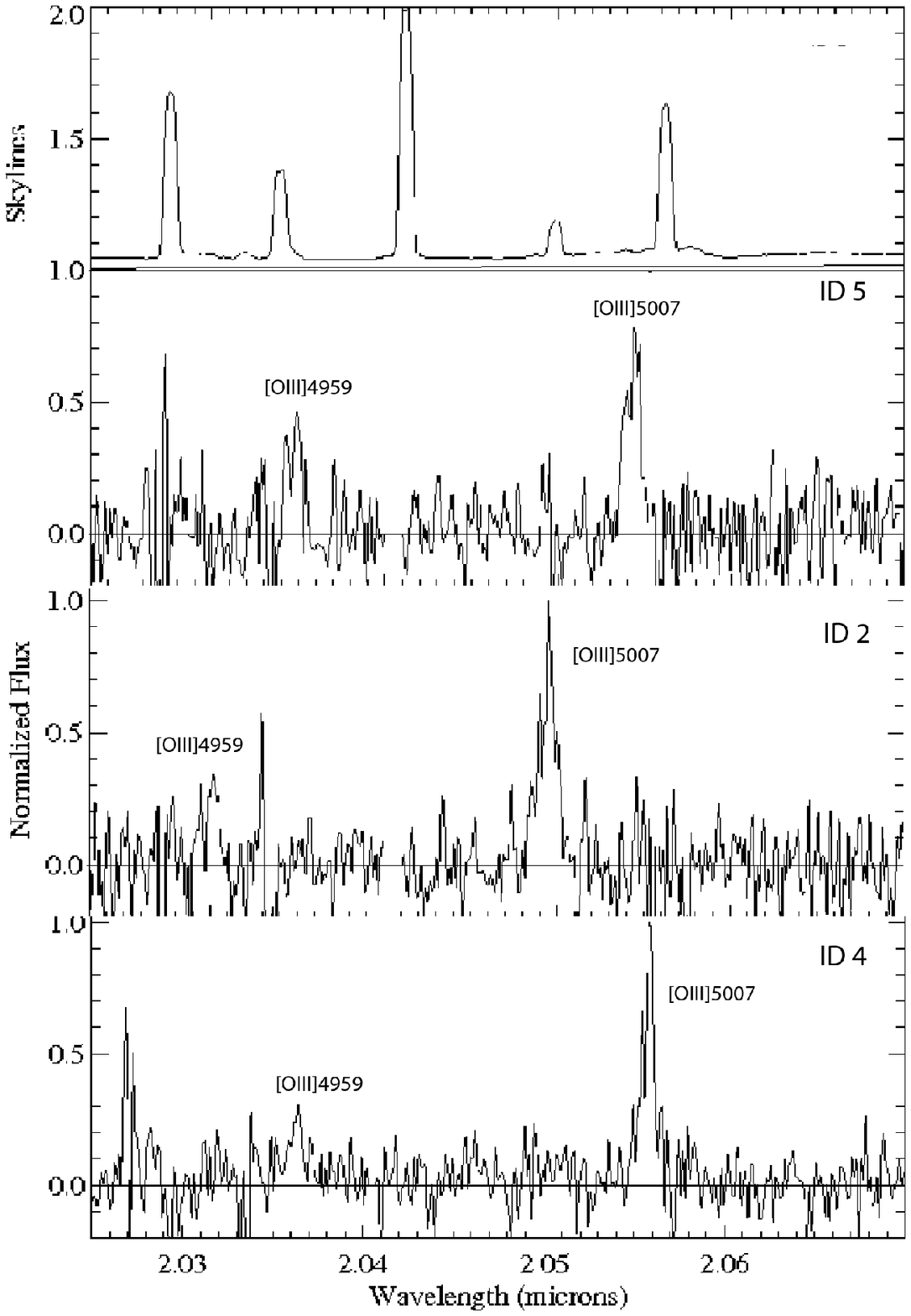}
\caption{One--dimensional spectra of the three confirmed [O{\sc iii}] emitters together with the night emission skylines. The [O{\sc iii}]$\lambda$ $\lambda$5007,4959 doublet lines are both clearly visible in each spectrum. A residual emission line feature at 2.04 $\mu$m has been mashed out. \label{spectra}}
\end{figure}

The spectra of the [O{\sc iii}] galaxies are presented in
Fig.~\ref{spectra} together with the night sky emission line
spectrum. Three of the four targets (ID 2, 4, and 5) exhibit a strong [O{\sc
    iii}]$\lambda$5007,4959 emission line doublet shifted into the window 3.095 $ 
\le $ \textit{z} $ \le $ 3.105. The presence of both lines of the [O{\sc
iii}]$\lambda$ $\lambda$5007,4959 doublet in these spectra confirm that the selected targets are [O{\sc
iii}] emitters. Table \ref{oiii_detections} lists the redshifts of the three confirmed [O{\sc iii}] emitters, the radio
galaxy MRC\,0316-257 and the four [O{\sc iii}] candidates, that are
also Ly$\alpha$ emitting galaxies spectroscopically confirmed to be members of the MRC\,0316-257 protocluster. Fig.~3 shows the redshift distribution of the spectroscopically confirmed [O{\sc iii}] emitting galaxies together
with the redshift distribution of the spectroscopically confirmed
Ly$\alpha$ emitting galaxies that comprise part of the protocluster around
MRC\,0316-257. The redshifts of the Ly$\alpha$ emitting galaxies are
taken from \citet{Venemans2005}. They were selected using a narrow-band filter sensitive to the redshift range 3.12--3.17.\\
The three spectroscopically confirmed [O{\sc iii}] emitting
galaxies occupy a narrow range in velocity distribution, although their redshifts falls $\sim$2100\,km\,s$^{-1}$ blueward of the mean redshift of the MRC\,0316-257 protocluster.

\begin{figure}
\includegraphics[width=1\columnwidth]{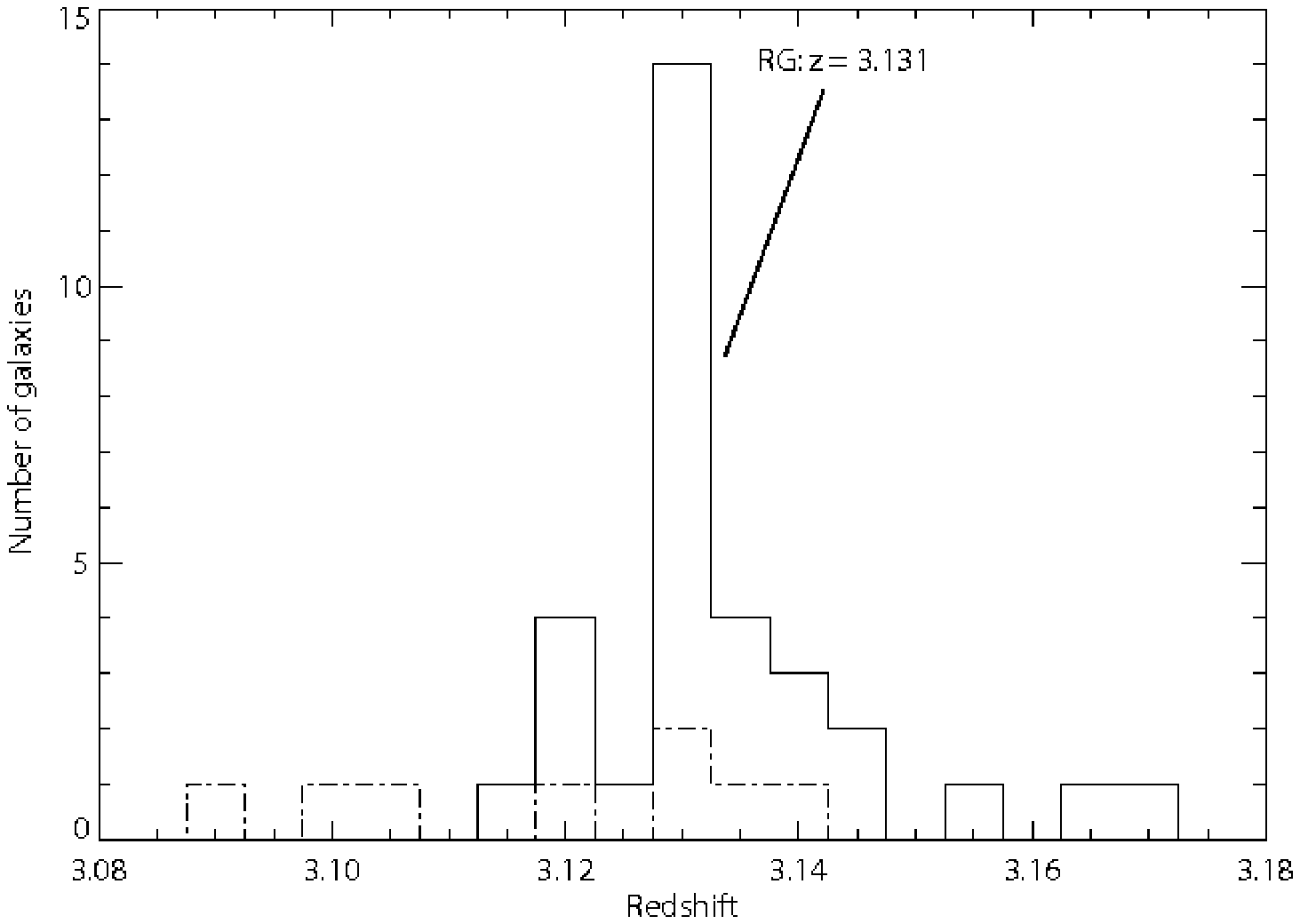}
\caption{Redshift histogram of the spectroscopically confirmed [O{\sc iii}] emitters (dash-dotted line) together with the previously known 32 Ly$\alpha$ emitters (solid line). Ly$\alpha$ emitters also include the radio galaxy. \label{histogram}}
\end{figure}

Although one candidate [O{\sc iii}] emitter (ID 9) was not detected, this non-detection is not significant. The object is almost a factor of four fainter than the faintest spectroscopically detected galaxy and its narrow--band flux would only have resulted in a  marginal detection in the available spectroscopic exposure time.  

\begin{table*}
 \centering 
\begin{tabular}{|r|c|c|c|c|c|c|c|}
  \hline
  Source ID & RA (J2000) & Dec (J2000) & [O{\sc iii}] flux & Ly$\alpha$ flux & Redshift & m$_{\rm V}$ & comments \\
 &  &  & $10^{-17} $erg\,s$^{-1}$cm$^2$ & $10^{-17} $erg\,s$^{-1}$cm$^2$ &  &  & \\
(1) & (2) & (3) & (4) & (5) & (6) & (7) & (8)\\
   \hline 
1	&  03:18:12.1  & --25:35:10.5 &  169 $\pm1.7$  & 84   & 3.131$_{Ly \alpha}$ & 23.8 & RG, LBG, DRG, LAE\\
2	&  03:18:08.6  & --25:35:28.8 &  39.0$\pm5.1$ & --    & 3.094$_{[O III]}$ & 25.3 & LBG\\
3       &  03:18:09.0  & --25:34:59.3 &	 8.7 $\pm1.0$ & 18.0  & 3.136$_{Ly \alpha}$ & 24.7 & LBG, DRG, LAE\\
4	&  03:18:15.1  & --25:36:37.8 &  8.5 $\pm0.9$ & --    & 3.106$_{[O III]}$ & 25.7 & LBG \\
5	&  03:18:06.6  & --25:35:20.1 &	 8.3 $\pm1.3$ & --    & 3.104$_{[O III]}$ & 26.4 & DRG\\
6       &  03:18:12.6  & --25:36:23.0 &  5.9 $\pm0.9$ & 4.6   & 3.123$_{Ly \alpha}$ & --   & LAE\\
7	&  03:18:16.8  & --25:35:46.1 &  4.3 $\pm0.9$ & 6.3   & 3.131$_{Ly \alpha}$ & 24.9 & LBG, LAE\\
8       &  03:18:13.7  & --25:35:07.7 &  4.2 $\pm0.8$ & --    & --    & 23.9 & LBG\\
9       &  03:18:12.7  & --25:36:44.4 &	 2.4 $\pm0.7$ & --    & --    & 26.9 & --\\
10	&  03:18:08.8  & --25:35:08.3 &	 2.3 $\pm0.5$ & --    & --    & 25.7 & LBG\\
11	&  03:18:12.0  & --25:34:52.9 &	 2.3 $\pm0.6$ & 1.5   & 3.141$_{Ly \alpha}$ & 26.5 & LAE\\
12	&  03:18:15.3  & --25:35:41.1 &	 1.5 $\pm0.4$ & --    &  --   & --   & --\\
13	&  03:18:16.1  & --25:36:19.9 &	 1.3 $\pm0.4$ & --    &  --   & 25.3 & --\\
\hline 
14      &  03:18:09.7  & --25:36:47.5 &     $<$1.1    & 2.9   & 3.124 &  --  & LAE \\
15      &  03:18:14.1  & --25:35:54.3 &     $<$1.2    & 1.1   & 3.132 &  --  & LAE \\
16      &  03:18:15.2  & --25:35:28.9 &     $<$0.5    & 1.0   & 3.146 &  --  & LAE \\
17	&  03:18:06.9  & --25:35:26.2 &     $<$0.7    & 0.8   & 3.143 &  --  & LAE \\
18      &  03:18:15.1  & --25:35:13.1 &     $<$0.7    & 0.72  & 3.131 &  --  & LAE \\
     
\hline
\end{tabular}
\caption{Properties of the [O{\sc iii}] emitters in the field together with confirmed Ly$\alpha$ emitters within the ISAAC field-of-view. Column 1 gives the assigned source numbers and columns 2 and 3 the equatorial celestial coordinates  respectively right ascension and declination. Columns 4 and 5 give the [O{\sc iii}] and Ly$\alpha$ fluxes in units of $10^{-17} $erg\,s$^{-1}$cm$^2$, errors are at 1$\sigma$. Ly$\alpha$ fluxes are determined from same apertures as [O{\sc iii}] fluxes. The [O{\sc iii}] flux given in column 4 for these galaxies is a 2$\sigma$ upper limit. The redshifts are given in column 6 and the line used to obtain them is specified as subscript. The visual magnitudes are shown in column 7. Column 8 gives the comments about the galaxy types where the label RG indicates the radio galaxy, LAE indicates the galaxy is a spectroscopically confirmed Ly$\alpha$ emitting galaxy in the protocluster, DRG indicates the galaxy is a candidate ''Distant Red Galaxy", and the label LBG indicates that the galaxy is a candidate Lyman break galaxy (LBG) at the redshift of the protocluster.
Galaxy ID 3 has extended Ly$\alpha$ and [O{\sc iii}] emission. Galaxies ID 14-18 are not detected in [O{\sc iii}] but are within the ISAAC field-of-view and are spectroscopically confirmed Ly$\alpha$ emitters within the protocluster. Photometry of the radio galaxy is unreliable as it lies very close to a foreground galaxy.  \label{oiii_detections}}
\end{table*}

\subsection{Extinction--corrected Star Formation Rates of the [O{\sc iii}] Emission Line Galaxies}

\subsubsection{Star Formation Rates from UV Continuum}

Under the assumption that the galaxies are dominated by hot, young massive stars and are relatively dust free, one can use the luminosity of the UV continuum to estimate a star formation rate (SFR).
We convert the r$_{625}$ magnitudes of the [O{\sc iii}] candidates into a SFR using

\begin{eqnarray}
SFR[M_{\odot}~yr^{-1}] & = & \frac{4 \pi D_{L}^2 F_{\nu}}{(1+z) 8\cdot10^{27}[erg~cm^{-2}~s^{-1}~Hz^{-1}]} 
                                                                                                \nonumber \\
                       & = & \frac{L(1500 \AA)}{8\cdot10^{27}[erg~s^{-1}~Hz^{-1}]} 
\label{starformationrate}
\end{eqnarray} 

\noindent for a Salpeter IMF and where $D_{L}^2$ is the luminosity distance and $F_{\nu}=10^{(-0.4\cdot(r_{AB}+48.6))}$ the flux density \citep{Madau1998}.
We determine a reddening correction from the UV continuum slope, derived from the measured $r_{625}-I_{814}$ colour. 
We assume the galaxies have a standard power-law spectrum with slope $\beta$ ($F_{\lambda} \propto \lambda^{\beta})$, so that a spectrum that is flat in $F_{\nu}$ has $\beta = -2$. We define

\begin{equation}
\beta_{rI}=-0.4\times \frac{(r_{625}-I_{814})}{log_{10}\frac{\lambda_r}{\lambda_I}}-2 
\label{uv_slope}
\end{equation}  

\noindent Where $\lambda_r$ and $\lambda_I$ are the central wavelengths of the r$_{625}$ and I$_{814}$ ACS-filters. In general, we note that the UV slope $\beta$ gives a reasonable estimate for the extinction 
if the object has a 'pure' starburst spectrum and is not a mix of old and new populations \citep{Kong2004}..
We determine the extinction through 

\begin{equation}
E(B-V)=\frac{\beta-\beta_0}{8.067},
\label{extinction1}
\end{equation}

\noindent with $\beta_0=-2.5$, as expected for ionizing populations \citep{Meurer1995}. The $\beta_0$ changes somewhat as a function of age and which part of the UV spectrum is being used to estimate the slope. Meurer \& Heckman (1995)  derived $\beta_0$ from a fit to the UV spectrum over the wavelength range that we approximate here with r$_{625}$-I$_{814}$. A wide range of star formation histories, from short bursts less then 10 Myr old to continuous star formation 100 Myr and older will produce a spectrum with $\beta_0 \approx -2.5$. The range in $\beta_0=-2.5$ of $\pm0.2$ for ionizing populations amounts to an uncertainty in obscuration $A(1500)$ of $\pm0.21$ mag. We derive the obscuration at 1500 $\AA$ ($A(1500)$) using the correlation in \citet{Calzetti2000}:

\begin{equation}
A(1500)=4.39E(B-V),
\label{obscur1}
\end{equation}

\noindent although  we note that this correlation has a relative large scatter. 
We present the extinction-corrected UV SFRs in Table 2.

We do not include the UV continuum and [O{\sc iii}] SFRs for the radio galaxy because the photometry is unreliable as it lies very close to a luminous foreground galaxy.
However, we note that the UV continuum of the radio galaxy is dominated by emission from stars. The contribution from scattered AGN light is limited by the upper limit on the polarization of the continuum of $P<4$ \% \citep{DeBreuck2005}. If all the light at a rest--frame of $1500 \rm{\AA}$ is due to young stars, the SFR of the radio galaxy is $\sim100 M_{\odot} yr^{-1}$, (uncorrected for dust absorption). This is similar to the uncorrected SFRs in radio  galaxies at $z\sim2.5$ \citep{Vernet2001} as calculated from the rest--frame UV continuum. However, the calculated SFR is likely to be a lower limit because radio galaxies are known to have significant amounts of obscured star formation (Stevens et al 2003).

\subsubsection{Star Formation Rates from [O{\sc iii}] Emission}
\citet{Kennicutt1992} demonstrated that the large dispersion in the [O{\sc iii}]/H$\alpha$ ratio among star forming galaxies makes [O{\sc iii}] an unsuitable emission line for determining star formation rates (SFRs).
The large dispersion is dominated by the variation in excitation mechanism and
oxygen abundance, as well as reddening
\citep{Moustakas2006}.
However, \citet{Moustakas2006} note that [O{\sc iii}] can be used to make a crude estimate of the SFR  and in particular can be used to estimate minimum SFRs. Fig.~13 from \citet{Moustakas2006} shows that there is a maximum [O{\sc iii}]/H$\alpha$ ratio observed from star forming galaxies which is $\sim2.4$ \citep{Teplitz2000}. Using this maximum ratio and the conversion between H$\alpha$ and SFR from \citet{Kennicutt1998}:
\begin{equation}
SFR_{\rm H\alpha}~[{\rm M_{\odot} ~ yr^{-1}}]=7.9\times 10^{-42}\frac{\rm L_{\rm H\alpha}}{\rm erg ~ s^{-1}},
\label{sfrha}
\end{equation}
we derived a lower limit to the SFR from the [O{\sc iii}] luminosity to be: 

\begin{table*}
 \centering 
\begin{tabular}{|r|c|c|c|c|}
  \hline
  Source ID & Corr-SFR(UV[1550$\AA$]) & Corr--SFR ([O{\sc iii}]) & $\frac{SFR ([O{\sc iii}])}{SFR (UV)}$ & E(B-V)$_{\rm UV}$\\
 & [M$\odot$/yr] & [M$\odot$/yr] &  & \\
(1) & (2) & (3) & (4) & (5)\\
   \hline 
1	& 439.8& --    & --   & 0.358 $\pm0.007$\\
2	& 9.1  & 252.8 & 27.7 & 0.090 $\pm0.032$\\
3       & 17.1 & 44.5  & 2.6  & 0.061 $\pm0.036$\\
4	& 15.3 & 122.2 & 8.0  & 0.173 $\pm0.034$\\
5	& 5.1  & 36.1  & 7.1  & 0.045 $\pm0.088$\\
6       & 1.4  & 63.3  & 45.2 & 0.141 $\pm0.065$\\
7	& 16.5 & 22.1  & 1.4  & 0.065 $\pm0.022$\\
8       & 150.1& 45.1  & 0.3  & 0.136 $\pm0.065$\\
9       & 5.1  & 20.4  & 4.0  & 0.124 $\pm0.100$\\
10	& 8.0  & 14.6  & 1.8  & 0.084 $\pm0.048$\\
11	& 5.1  & 36.2  & 7.1  & 0.180 $\pm0.062$\\
12	& 2.4  & 7.3   & 3.1  & 0.055 $\pm0.115$\\
13	& 16.7 & 20.7  & 1.2  & 0.184 $\pm0.034$\\

\hline
\end{tabular}
\caption{Dust--corrected UV- and [O{\sc iii}]-SFRs of the 13 [O{\sc iii}] emitters in the field. Column 1 gives the source IDs. The UV-SFRs and the [O{\sc iii}]-SFRs of the emitters, both corrected for dust obscuration, are shown in columns 2 and 3 respectively. SFRs are given as M$\odot$/yr. Column 4 gives the corrected $\frac{SFR [O{\sc iii}]}{SFR [UV]}$ ratios and column 5 the UV extinctions for all candidates. Candidate 14--18 (see Table 1) which were not detected in [O{\sc iii}], are not displayed. \label{acs_detections}}
\end{table*}

\begin{equation}
SFR_{\rm [O{\sc iii}]}~[{\rm M_{\odot} ~ yr^{-1}}]>0.33\times 10^{-41}\frac{\rm L_{\rm [O{\sc iii}]}}{\rm erg ~ s^{-1}}.
\end{equation}
We note that the extinction derived from the ionized gas (Fanelli et al. 1988) is not the same as the extinction derived from the UV stellar continuum \citep{Calzetti1999, Calzetti2000}.
We have therefore used the calibration from \citet{Calzetti2000} linking the stellar continuum colour excess to the colour excess derived from the nebular gas emission lines:

\begin{equation}
E(B-V)_{stars}=(0.44 \pm 0.03)E(B-V)_{gas}.
\label{gas extinction}
\end{equation}

\noindent We determined  the intrinsic [O{\sc iii}] line flux through

\begin{equation}
F_i(\lambda)=F_o(\lambda)10^{0.4E(B-V)_{gas}k'(\lambda)}
\label{Intrinsic flux}
\end{equation}

\noindent where $F_i(\lambda)$ and $F_o(\lambda)$ are the intrinsic and observed [O{\sc iii}] line flux densities, respectively, and $k'(\lambda)$ is defined as $k'(\lambda)=A'(\lambda)/E(B-V)_{gas}$ and given by:

\begin{eqnarray}
k'(\lambda)& = & 2.659(-2.156+1.509/\lambda-0.198/\lambda^2+0.011/\lambda^3)
                                                         \nonumber \\
	   & + & R'_V,
                                                         \nonumber \\
           &   &  0.12\mu m \le \lambda \le 0.63 \mu m
\label{reddening curve}
\end{eqnarray}

\noindent where $\lambda$ is 0.5007 $\mu m$ the rest--frame wavelength of the [O{\sc iii}] emission line, and the default value of $R'_V$ is 4.05 based on comparison with starburst galaxies \citep{Calzetti1999, Calzetti2000}.

\begin{figure*}
\includegraphics[width=0.7\textwidth]{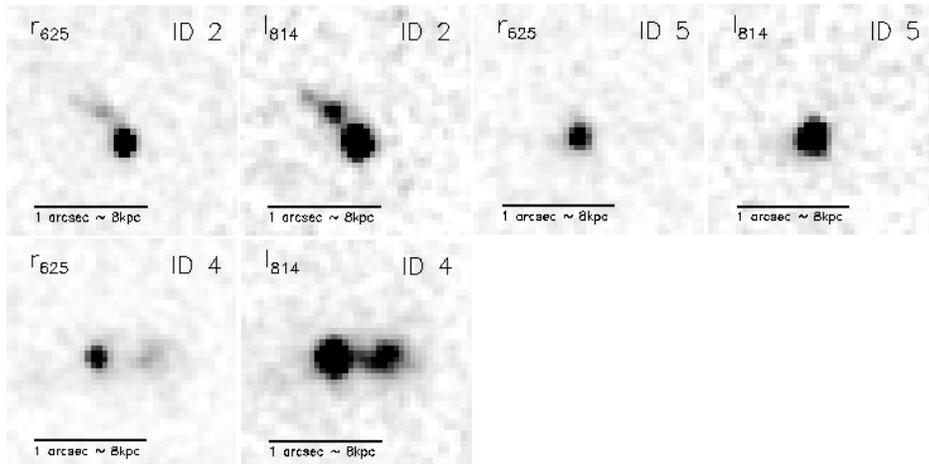}
\caption{ACS r$_{625}$-band (left columns) and I$_{814}$-band (right columns) images of the three confirmed [O{\sc iii}] emitters. From top to bottom: ID 2, 4 and 5. The scale is 0.05 arcsec pixel$^{-1}$ and each box is 2" on each side. Only ID 5 has a single component in both UV bands. The other two objects instead reveal, most clearly in the I-band, the presence of a secondary ``core'' (ID 2) or of an extended structure (ID 4). The derived corrected SFR for object 2 indicates that it could be an obsured AGN. \label{postage}}
\end{figure*}

We list extinction-corrected [O{\sc iii}] SFRs  and the ratio $\frac{SFR [O{\sc iii}]}{SFR [UV]}$ in Table 1.
This ratio shows that approximately half (7/13) of the candidates have a comparable SFR derived from the UV continuum and [O{\sc iii}] emission, whilst 3/13 have a ratio of approximately 7--8, and 2/13 have very high $\frac{SFR [O{\sc iii}]}{SFR [UV]}$ ratios. A large discrepancy between the derived SFRs on the basis of UV continuum and [O{\sc iii}] emission, i.e. a high $\frac{SFR [O{\sc iii}]}{SFR [UV]}$, may indicate a different or additional [O{\sc iii}] excitation source other than star formation.
We note that the [O{\sc iii}] SFRs are lower limits as we used the maximum [O{\sc iii}]/H$\alpha$ ratio of 2.4. Hence the $\frac{SFR [O{\sc iii}]}{SFR [UV]}$ may be larger by this factor. \citet{Moustakas2006} show that variations in chemical abundance, ionization parameter 
and reddening all increase the scatter in the
relation between the SFR as derived by H$\alpha$ emission and [O{\sc
    iii}] luminosity. They conclude that the $1\sigma$ uncertainity
associated with converting the [O{\sc iii}] luminosity into a SFR is a
factor of 3-4. We conclude that the [O{\sc iii}] luminosity of only
two galaxies (ID 2 and 6) is beyond that expected from its UV-determined star
formation with more than $3\sigma$ confidence, therefore they require an additional excitation source --most likely an AGN. We don't quote measurement errors for the SFRs in the Tables because they are much smaller than the intrinsic scatter of 3-4 in the conversion.

\begin{figure*}
\includegraphics[width=0.65\textwidth]{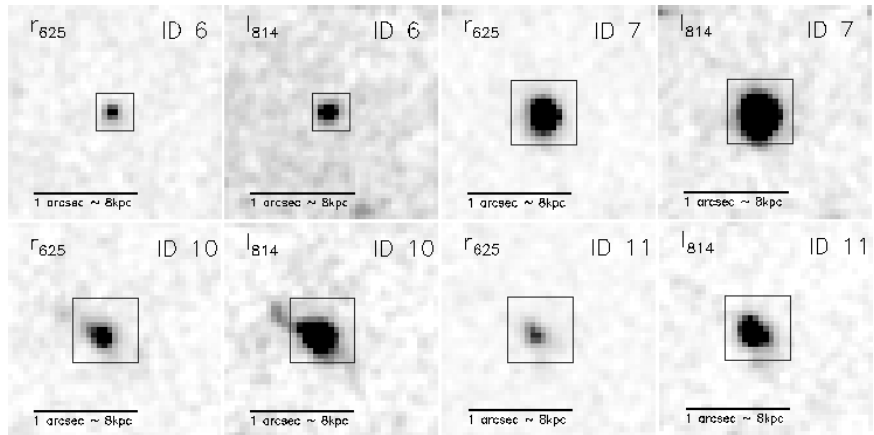}
\caption{ACS r$_{625}$-band (left column) and I$_{814}$-band (right column) images of the 4 ``compact'' [O{\sc iii}] candidates. The scale is 0.05 arcsec pixel$^{-1}$ and each box is 2" on each side. The detection squares are centered on the detection pixel and their sides match the sizes of the objects (see Sec. 3.3 and Table 3 for the values).Consistently with the mean half light radius of unsaturated stars in the ACS field of $\sim0\farcs09$ we cannot exclude the presence of unresolved components within these "compact" objects. \label{candidates1}}
\end{figure*}

\begin{figure*}
\includegraphics[width=0.65\textwidth]{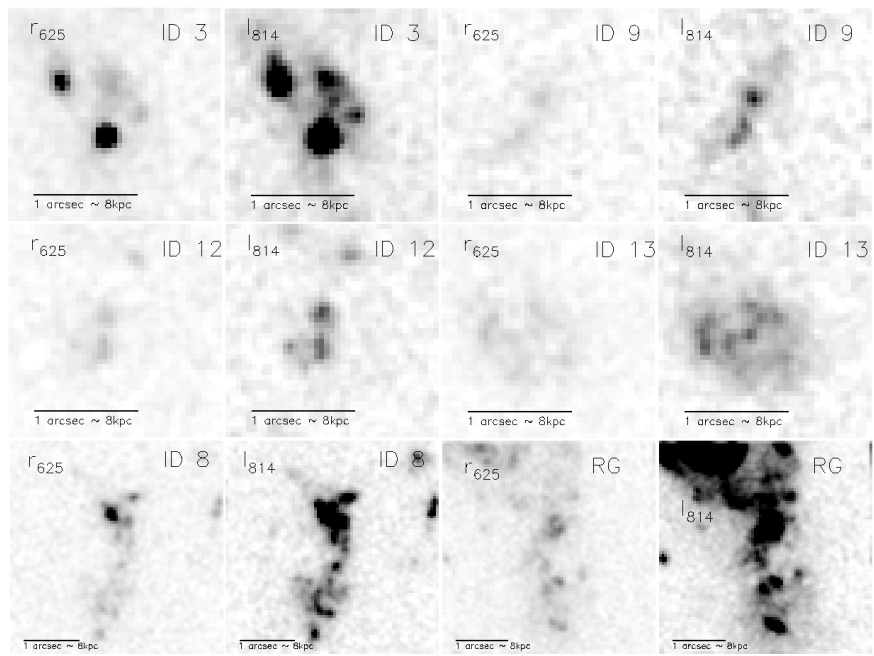}
\caption{ACS r$_{625}$-band (first and third columns) and I$_{814}$-band (second and fourth columns) images of the 6 "diffuse and clumpy" objects. The scale is 0.05 arcsec pixel$^{-1}$ and each box is 2" on each side. The radio galaxy is showed together with object ID 8 (the probable merger) at the bottom row. The dimension of their detection boxes is 4". Objects that have larger sizes and are diffuse or clumpy are plotted without the detection squares.\label{candidates2}}
\end{figure*}

\subsection{Sizes and Morphologies of the [O{\sc iii}] Emitters }
Figures 4, 5 and 6 show r$_{625}$ and I$_{814}$ ''postage-stamp" regions of the 13 [O{\sc iii}] candidates. From a visual inspection approximately half (8 objects) of the [O{\sc iii}] candidate emitters have morphologies characterized as a bright compact core surrounded by diffuse emission with significantly lower surface brightness. The remaining 5 candidates (including the radio galaxy shown in Fig. 6) have diffuse or clumpy morphologies. These objects have larger sizes and do not posses an obvious core.
To quantify the sizes of these objects, the half light radius ($r_H$) of each emitter was measured using the program SExtractor on the I$_{814}$-band images. The half light radius is defined as the radius of a circular aperture in which the flux is 50\% of the total flux.
The half light radii of the candidates range from 0.06 to 0.75 arcsec.
The sizes of the candidate emitters in the $I_{814}$--band are given in Table 3 . Galaxies that are diffuse or clumpy have larger uncertainties due to their irregular shapes.  The measured physical sizes correspond to 1.2--4.8 kpc. 
The mean half light radius of isolated, unsaturated stars in the ACS field was found to be $\sim0.06$ arcsec. One of the [O{\sc iii}] candidates, ID 6, is classified as unresolved. We compared the measured sizes of the confirmed Ly$\alpha$ emitters (IDs 3, 6, 7 and 11) with that of \citet{Venemans2005}. The sizes of the confirmed Ly$\alpha$ emitting sources in \citet{Venemans2005} were also derived from the analysis of an ACS image taken within the I$_814$ filter, and the halflight radius of each emitter was measured using the program SExtractor too.  We found a good match within an error of $\sim0.5$ kpc.
The mean half-light radius of the [O{\sc iii}] emitters is 1.8 kpc, which is comparable to that of luminous LBGs at z$\sim3$ \citep{Bouwens2004, Ferguson2004}.

We find a range of morphologies from simple compact to clumpy and diffuse typical of LBGs \citep{giavalisco1996, Lotz2004, Lotz2006, Elmegreen2007}.
The ACS image of the radio galaxy shows several objects within 3\arcsec ($\sim 25$ kpc), surrounded by low surface brightness emission ($\ge 24.8$ mag arcsec$^{-2}$). Such a structure is typical for the rest-frame UV emission from the hosts of the most luminous, radio-loud sources at high redshift \citep{Pentericci1999, Venemans2005, Zirm2005, Miley2006, Hatch2008}. 

Objects IDs 3, 8, 9, 12 and 13 are clumpy with two or three clumps clearly visible in the I$_{814}$-band image (Fig. 6). They are probably single extended objects with a lot of structure and clumps that we interpret as different regions of star formation, perhaps induced by a galaxy merger. According to Elmgreen et al. (2005) the clumpy objects may not be mergers,but rather unstable discs.
Objects IDs 2, 4, 5 and IDs 6, 7, 10 and 11 are identified with single objects in the ACS images (Fig. 4 and 5).
However, in some cases where the objects appear to be single ones in the [O{\sc iii}], the I$_{814}$ images reveal ``double-core'' (ID 2, 4, 9 and 10). On average those two cores are separated less than 0.5 kpc from each other.

\begin{table}
 \centering 
\begin{tabular}{|l|c|c|c|}
  \hline
  Source ID & $r_h$ (arcsec.) & $r_h$ (kpc)& morphology \\
 & [arcsec] & [kpc] & \\
(1) & (2) & (3) & (4)\\
   \hline
1	& -- & -- & clumpy (RG)\\   
2	& 0.13 & $\sim 1.0$ & double--core \\ 	   
3*	& 0.26 & $\sim 2.0$ & clumpy \\
4	& 0.27 & $\sim 2.1$ & double--core \\
5	& 0.10 & $\sim 0.8$ & single compact--core \\
6*	& 0.06 & $  <0.5  $ & single compact--core \\
7*	& 0.20 & $\sim 1.5$ & single compact--core  \\
8	& 0.75 & $\sim 5.8$ & clumpy \\
9	& 0.15 & $\sim 1.2$ & single--core \\
10	& 0.17 & $\sim 1.3$ & single compact--core\\
11*	& 0.12 & $\sim 0.9$ & single compact--core\\
12	& 0.19 & $\sim 1.5$ & clumpy \\
13	& 0.43 & $\sim 3.3$ & clumpy \\
\hline
\end{tabular}
\caption{Sizes and morphologies of the [O{\sc iii}] candidate emitters located within the field of ACS. Column 1 gives the source IDs. The half-light radii in arcsec and Kpc are given in column 2 and column 3 respectively. Column 4 gives information about the morphology of the emitters. The sizes of the galaxies which are diffuse or are multiple objects have a larger uncertainty in their values because of the irregular shapes. The (*) labels the confirmed Ly$\alpha$ emitters from \citet{Venemans2005}. ID 6, having a $r_h \sim 0\farcs06$, is unresolved. \label{acs_sizes}}
\end{table}

We note that the compact clumps seen in the HST images may contain substructure on a smaller scale than the HST resolution, that prevents one from concluding that they have spheroiodal or disky morphologies. This can be seen from a comparison with the rest-frame UV morphologies of a sample of local analogs of high redshift LBGs as well as with gravitationally lensed LBGs \citep{Overzier2007}.
Both clumpy and compact core-dominated galaxies clearly show extended diffuse emission around or extending from them, which could be an indication of interacting systems \citep{Younger2007}. 
In a few cases, the [O{\sc iii}] emission coincides with the mean centroid of the UV continuum. In other cases, the [O{\sc iii}] emission mainly comes from one of the two components visible in the UV. 
Fig. 6 shows three objects (IDs 3, 9 and 12) that associate the position of the single [O{\sc iii}] selected candidate with respectively a double (IDs 3 and 9) and a triple (ID 12) detection in the ACS UV images (most clearly seen in the I$_{814}$). In all the three cases the detection box gives the position of the candidate as referred in the [O{\sc iii}] selected list.
In summary, the [O{\sc iii}]-emitting galaxies show a range of diverse morphologies.

\section{discussion}
\subsection{Space Density of [O{\sc iii}] Emitting Galaxies}
\noindent
How do the redshifts of galaxies detected on the basis of their
redshifted [O{\sc iii}] emission relate to the protocluster at $z=3.13$
found by Venemans et al. (2005) ? At $z>2.8$, [O{\sc iii}]$\lambda5007$ is
the most easily observed bright emission line in star-forming galaxies, since
H$\alpha$ is redshifted beyond the near-infrared and Ly$\alpha$ is only
present in a small fraction of LBGs (Pettini et al. 1998, Shapley et
al. 2003).  However, the selection of [O{\sc iii}] emitters at high redshift
using narrow-band imaging in the infrared is far less common than
similar techniques used to target Ly$\alpha$ or H$\alpha$ \citep{Teplitz2000, Moorwood2000}.  Consequently, not much is
known about the number density and clustering statistics of
[O{\sc iii}]-selected objects at $z\sim3$.

\citet{Moorwood2000} detected five [O{\sc iii}] emitters at redshift 3.1
in a 18.9 arcmin$^2$ field with flux greater than $8\times10^{-17}$
erg cm$^{-2}$ s$^{-1}$, consistent with the statistics of 1 object in 3 arcmin$^2$ found
by Teplitz et al. (1999). Including the radio galaxy, we detect 13
candidate [O{\sc iii}] emitters in this field, and we have obtained
redshifts for 7 of these. Four objects had previously determined redshifts
based on Ly$\alpha$, and are among the protocluster galaxies at
$z\approx3.13$. The remaining three objects were spectroscopically
identified as galaxies at $z\approx3.10$, and have colours consistent
with Lyman break galaxies (objects \#2 and \#4 in Table 1) and red galaxies (\#5).  It thus seems
likely that the remaining [O{\sc iii}] emitters also have redshifts $z\sim3.1$. 
If we constrain our sample to the same limiting depth as obtained by
Moorwood et al. (2000), and take into account that our field area is
3.5 times smaller, we obtain a local density in the [O{\sc iii}] field of 3.5 x the field density and finally, according eq. 21 in \citet{Venemans2005}, a galaxy overdensity of 2.5.
If the [O{\sc iii}] emitting galaxies are a subset of the general population
of star-forming galaxies at $z\sim3$, their
distribution can be expected not to be random, but relatively strongly
clustered both in real space and in velocity space (e.g. Ouchi et
al. 2004, Monaco et al. 2005, Lee et al. 2006). This makes it
difficult to assess the statistical significance of the apparent
enhancement in the observed number density of [O{\sc iii}] emission line galaxies, 
without additional observations or a detailed comparison
with simulations.  Nevertheless, the observed space density of [O{\sc iii}] galaxies is consistent
with the the overdensity of 2.3 derived for Ly$\alpha$ galaxies in the MRC\,0316-257 protocluster by 
Venemans et
al. 2005).

It is interesting to note that the three newly obtained redshifts all
have $z\approx3.10$, compared with $z\approx3.13$ for the Ly$\alpha$ protocluster.
These [O{\sc iii}] emitters appear therefore to be 
blueshifted by 2100 km s$^{-1}$ relative to the radio galaxy and its
associated protocluster of Ly$\alpha$ emitters.  It is unclear whether this
apparent shift truly reflects the distribution of the [O{\sc iii}] emitters
in this field, or whether it is due to small-number statistics
resulting from the $\sim50$\% incompleteness of our spectroscopic
follow-up.\\
It is unlikely that the [O{\sc iii}] lines are systematically shifted with respect to the Ly$\alpha$ line, as the shift of 2100 kms$^{-1}$ is larger than the typical 600 kms$^{-1}$ shift that one would expect.
The [O{\sc iii}] filter covers a larger redshift range than the Ly$\alpha$ filter, extending to lower redshifts.
 It is possible that another structure exists at z$\sim$3.10 which is not detected through Ly$\alpha$ emitters due to the limited range of the Ly$\alpha$ filter.
The formal, co-moving distance between $z=3.13$ and
$z=3.10$ is $\sim30$ $h_{73}^{-1}$ Mpc.  Simulations of structure
formation show that protoclusters originate from gravitationally
collapsing regions as large as $\sim$ 20--40 Mpc at high redshift
(e.g. Suwa et al. 2006), and proto-cluster regions having sizes as
large as these have been found in a number of cases \citep[e.g.][]{Campos1999, Shimasaku2003, Hayashino2004, Intema2006}.
It is possible that the [O{\sc iii}] emitters at $z\approx3.10$ 
trace a sub-cluster that will eventually merge with the
''radio galaxy" protocluster at $z\approx3.13$. 
Alternatively, the numbers of 
galaxies in the vicinity of overdense regions may be significantly
enhanced relative to the general field due to an additional bias
associated with the larger walls or filaments in which proto-clusters
form. Such structures may have sizes of up to $\sim$100 Mpc on a
co-moving scale.

\subsection{Ionisation of [O{\sc iii}] Emission}

[O{\sc iii}] emission is commonly observed in the spectra of both star-forming and active galaxies. 
It is therefore interesting to consider whether the emission from our [O{\sc iii}] excess galaxies at 
$z\sim3.1$ is purely due to star-forming galaxies or whether there are contributions to their ionisation by 
AGN. The comparision of the [O{\sc iii}] and UV
continuum emission, and the morphology of the galaxies provide several
interesting clues as to the origin of the [O{\sc iii}] emission. At
low redshift the UV and optical continuum of star--forming galaxies
containing an AGN is dominated by the stellar light
\citep{Kauffmann2003}, while the emission lines are excited by both
the AGN and the young stars. Therefore objects for whose SFRs derived from [O{\sc
    iii}] emission are larger than those measured from their UV continua are likely 
to contain AGN that contribute to the ionisation of the O$^{+}$ gas. 

The SFRs derived from the [O{\sc iii}] luminosities are
comparable to the SFRs derived from the UV continua for approximately
half of the [O{\sc iii}] emitters. These [O{\sc iii}] galaxies
all have clumpy morphologies in the ACS images and their  [O{\sc iii}] emission is likely to be
ionised by
hot young stars.  The radio galaxy and ID 8 have very extended clumpy
structures that are suggestive of possible mergers or tidal
streams. Five [O{\sc iii}] galaxies have [O{\sc iii}] fluxes that are formally larger by a factor $>$ 7 compared with the fluxes expected from their UV-derived SFRs, indicating
that these galaxies harbour a hidden AGN.
All five of these galaxies have compact core--dominated morphologies and ID 2 has a nearby companion galaxy.
However, the uncertainties in the conversion between [O{\sc iii}] fluxes and SFR are large (factor of 3--4) and only in the cases of ID 2 and 6 are the enhanced [O{\sc iii}] luminosities significiant by greater than $3\sigma$.
If those two candidates are confirmed, the AGN fraction in this small field would be $\sim$14 \% of the total number of [O{\sc iii}] emitters compared with 5--10 \% for the AGN overdensity in Ly$\alpha$ emitting galaxies in protoclusters at $z\sim$2--4 \citep{Pentericci2002, Croft2005} and 3--5 \% for the AGN fraction in LBGs at $z\sim$3 \citep{Ouchi2007}.
Although the larger fraction of AGN in our [O{\sc iii}] small sample is only marginally significant, we note that this would be consistent with the fact that [O{\sc iii}] selection is known to be an efficient technique for finding low-luminosity AGN (type 2 Seyfert galaxies) in the optical \citep{Zakamska2004}. 
The conversion between SFR and [O{\sc iii}]
luminosity is too uncertain to be able to make any strong conclusions
as to the nature of the remaining 3 galaxies which have a larger
[O{\sc iii}] luminosity than can be explained by star formation given their UV SFRs.
We conclude that the [O{\sc iii}] galaxies detected at $z\approx3.10$ are ionized by a mixture of young stars and AGN.

\section{Conclusions and future work}
Searching for [O{\sc iii}]--emitting galaxies is a new feasible method for detecting protocluster members.  We have detected a new population of [O{\sc iii}] emitting galaxies in the neighbourhood of the radio galaxy MRC\,0316-257 at $z$=3.13. About half of the [O{\sc iii}] candidates
emitters are LBGs and a third were also previously detected by the Ly$\alpha$ selection technique. The [O{\sc iii}] technique complements narrow-band searches using
Ly$\alpha$ and H$\alpha$ emission, and observations of the Lyman and Balmer breaks, for finding members of protoclusters.
All of these different galaxy selection techniques are needed to study the different galaxy populations and to obtain a complete understanding of protocluster evolution.
13 candidate [O{\sc iii}] emitters were detected, including the radio galaxy, and 8 of these were spectroscopically confirmed. Three [O{\sc iii}] emitting galaxies lie in a small redshift interval at  $3.095<z<3.105$, which is blushifted by 2100\kmps\ with respect to the overdensity of Ly$\alpha$ emitting galaxies associated with the protocluster structure surrounding the radio galaxy.  These three [O{\sc iii}] emitters may trace a substructure that lies in front of the previously known 0316-257 protocluster. Further narrow-band imaging and spectroscopy of this field are necessary to establish the existence of such a structure.
A possible explanation for the [O{\sc iii}]--Ly$\alpha$ redshift separation are the difference in the filter band pass together with the existence of a possible big supersructure around the radio galaxy.
The observed space density of [O{\sc iii}] galaxies is consistent
with the overdensity of 3.3 derived for Ly$\alpha$ galaxies in the MRC\,0316-257 protocluster by \citet{Venemans2005}
The detected [O{\sc iii}] candidate emitters exhibit a wide range of morphologies. About half have compact structures with an unresolved core on the HST ACS image.  
All three spectroscopically confirmed [O{\sc iii}] emitters are compact. The median size of these objects is typically $\le 0\farcs24$, which at the average redshift $z=3.11$ correspond to $\sim 2$ kpc. 
The remaining sources are clumpy, consistent with merging or interacting systems. 
Using the next generation of wide-field infrared imagers and spectrographs, it
will be possible to increase the number of high-z detected [O{\sc iii}] emitters in the MRC\,0316-257 by an order of magnitude and carry out comprehensive infrared studies of several protoclusters at high redshift. Not only will such observations result in greatly improved statistics, but they will also facilitate detailed mapping of the cosmic web at high redshifts.

\section*{Acknowledgments}
This research is based on observations made with the VLT at ESO Paranal with program numbers 077.A-0310(A,B) and  078.A-0002(A,B), and on observations made with the NASA/ESA Hubble Space Telescope, obtained from the data archive at the Space Telescope Science Institute. STScI is operated by the Association of Universities for Research in Astronomy, Inc. under NASA contract NAS 5-26555.
NAH and GKM acknowledge funding from the Royal Netherlands Academy of Arts and Sciences.

\bibliographystyle{mn2e}\bibliography{mn-jour,o3}

\label{lastpage}

\end{document}